\documentclass[12pt,english]{iopart}
\usepackage{iopams}
\usepackage{setstack}
\usepackage{amstext}
\usepackage{babel}
\usepackage{graphicx}
\usepackage{amssymb}

\begin{document}


\title{Light scattering by a 
medium with a
spatially modulated optical conductivity: the case of graphene}

\author{ N. M. R. Peres$^1$, Aires Ferreira$^2$, Yu. V. Bludov$^1$, and M. I. Vasilevskiy$^1$}

\address{$^1$ Physics Department and CFUM, University of Minho, P-4710-057, Braga, Portugal.}

\address{$^2$Graphene Research Centre and Department of Physics, National University
of Singapore, 2 Science Drive 3, Singapore 117542}

\ead{peres@fisica.uminho.pt}

\date{\today}

\begin{abstract}
We describe light scattering from a graphene sheet 
having a modulated optical conductivity. 
We show that such modulation enables the excitation of surface plasmon-polaritons
by an electromagnetic wave impinging at normal incidence.
The resulting surface plasmon-polaritons are 
responsible for a substantial increase of electromagnetic radiation absorption by the 
graphene sheet. The origin of the modulation can be due  either to a 
periodic strain field or to adatoms (or absorbed molecules) with a 
modulated adsorption profile.
\end{abstract}

\pacs{81.05.ue,72.80.Vp,78.67.Wj}

\maketitle

\section{Introduction}

Since the days of A. Sommerfeld, back to 1899, that plasmonics effects in materials
and gratings are investigated both theoretically
and experimentally. Modern plasmonics gained a renewed interest with the discovery
super transmittance through a periodic array of holes with a size smaller than the 
diffraction limit \cite{Ebbesen1998}. The effect was explained invoking surface plasmons (SPs)
\cite{Ebbesen2003,Ebbesen2008,Stockman2011,PlasmonBook}. Beyond fundamental physics, plasmonics
encompasses wide range of applications, such as spectroscopy and sensing \cite{Amanda2005,Willets2007,Shalabney2011}, 
photovoltaics \cite{Green2011}, 
optical tweezers \cite{Reece2008,Juan2011}, 
nano-photonics \cite{Ozbay2006,Ebbesen2008}, transformation optics \cite{Ashkan}, etc.

A series of recent papers
 \cite{Schedin2010,LongJuPlasmonics,EchtermeyerPhoto} triggered
a burst of interest on plasmonic effects in graphene. In particular, it has been
shown that graphene has a strong plasmonic response in the THz frequency range 
at room temperature \cite{LongJuPlasmonics}. THz photonics is emerging as 
an active field of research \cite{Terahertz} 
and graphene may play a key role in THz metamaterials in the near future.
 Ju {\it et al.} \cite{LongJuPlasmonics} 
has shown that electromagnetic radiation impinging on a grid
of graphene micro-ribbons can excite SPs on graphene leading
to prominent  absorption peaks, whose position can be tuned by doping. 
In general, SPs cannot be excited  by directly  shining light in a
homogeneous system due to kinematic reasons: the momentum of a SP
 is much larger than that of the incoming light having the same frequency.
Therefore some type of mechanism is necessary to promote the excitation of
surface plasmons. The most common mechanisms for SP excitation are:
attenuated  total reflection (ATR) \cite{YuliyEPL}, 
scattering from a topological defect at the surface of the metal \cite{Ebbesen2008},
 and Bragg 
scattering using diffraction gratings (or producing a periodic corrugation) on the surface
of the conductor \cite{Toigo}. The method of Ju {\it et al.} is similar (but not exactly) to
patterning a metallic grating \cite{EchtermeyerPhoto} on top of graphene. 
A theoretical account of the experiment  by Ju {\it et al.} \cite{LongJuPlasmonics}
was given in a recent work \cite{Nikitin}. 

The reason why a periodic corrugation allows the excitation of SPs
can be understood by an analogy with the theory of electrons in a periodic potential.
The corrugation plays the role of the periodic potential. Therefore, the SP momentum 
 is conserved up to a reciprocal lattice vector, that is to say, 
the periodic corrugation provides the missing momentum needed to excite the 
SP. Another way of seeing the effect is to note that the grating gives rise to
a SP band structure in the first Brillouin zone. Then, the folding to 
plasmonic bands makes it possible to the impinging light to excite a SP
associated with the upper bands in the Brilloun zone. The excitation of 
SP modes in the first band is not possible in the grating configuration unless 
the ATR technique  is used.

The question we want to answer in this article is the following: 
in what circumstances can electromagnetic radiation (ER) 
impinging at the interface of a dielectric
and a metal excite SPs when the surface of the conductor
has no corrugation? Taking the example of graphene,
we show that a periodic modulation of the conductivity
suffices for that end. In situations where both modulation of the 
conductivity and corrugation are present, understanding the former
situation alone is a needed research step.

\section{Modulated conductivity}
\label{sec_mod_conductivity}

\begin{figure}[!h]
 \begin{center}
 \includegraphics*[width=10cm]{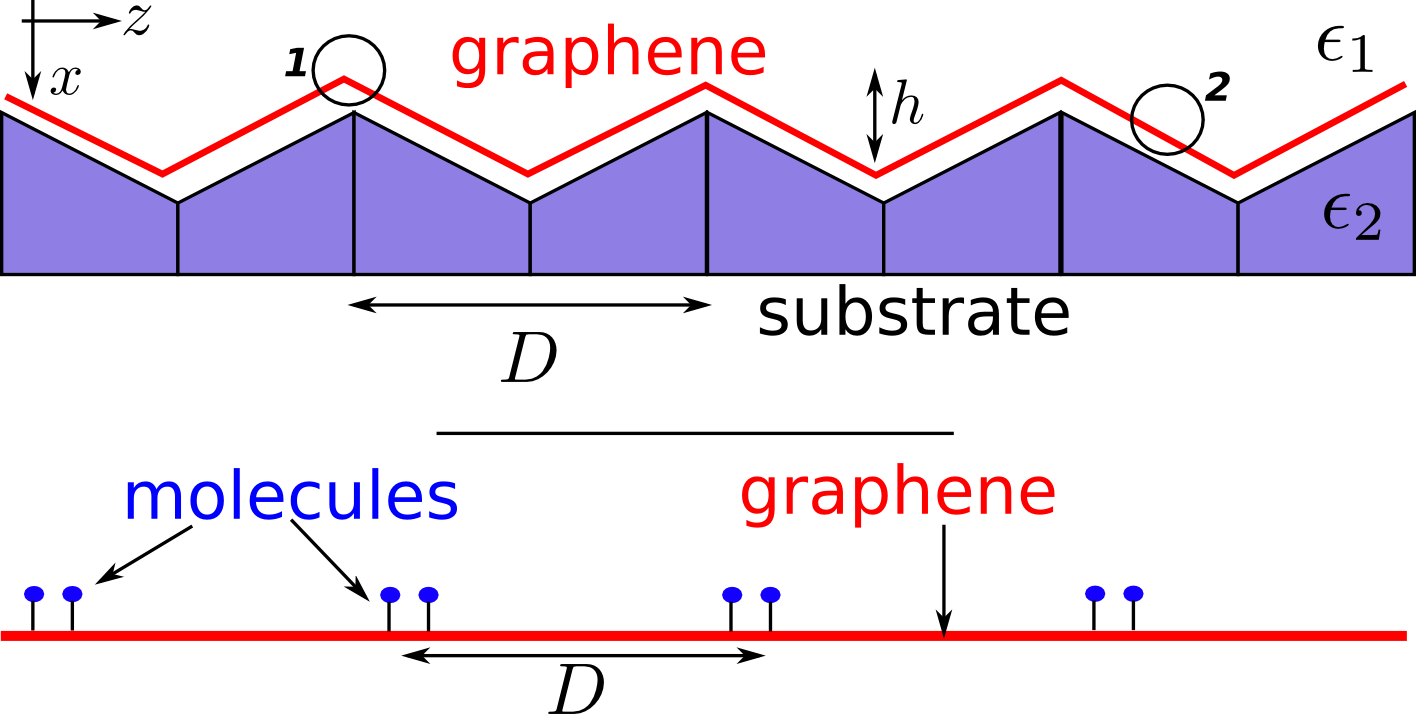}
 \end{center}
\caption{Periodic patterned substrate (top) and adsorbed atoms or molecules (bottom). 
For the patterned substrate we assume $h\ll D$ and, of course,
the graphene sheet will not follow the profile of the gate exactly.
The picture is meant to represent a possible away of inducing a periodic strain field
in graphene; the experimental realization may be quite different from this
representation. In the bottom figure we consider a different realization: 
the graphene sheet is locally doped by adsorbed atoms or molecules which 
create an inhomogeneous charge density profile and therefore a modulated
conductivity.}
\label{fig_saw_tooth}
\end{figure}

Several procedures can be used to induce a patterned conductivity on a 
graphene sheet.
One way is using split gates \cite{Davoyan2012}. 
This leads to a modulation in the electronic density
which induces a modulation of the conductivity.
One  can also assume that a CVD grown graphene sheet is transferred onto a patterned substrate, as
illustrated in Fig. \ref{fig_saw_tooth}. The transfer process together
with the patterning act in a way to produce inhomogeneous strain in the graphene
sheet. Qualitatively, we expect the strain to be higher in regions labeled
by 1 in Fig. \ref{fig_saw_tooth} than in those labeled by 2.
In a typical experiment we have
$h/D\sim 10^{-3}\ll1$. In this regime we can safely neglect the effect of the groves
in what concerns electromagnetic radiation scattering. Due to strain, the optical conductivity of graphene
will be spatially modulated with the period of the patterned substrate.
Naturally, the form of the modulated conductivity depends on the strain field.
It is a rather complex problem to determine the exact form of the spatial conductivity,
mainly because the strain field is not known exactly. In any case, we can say that the strain field
should have the same period of the patterned substrate.

Another possibility for producing a modulated conductivity is by molecular
adsorption. One can imagine an experimental procedure, for instance using a mask,
where graphene has regions exposed to molecular absorption separated from others
where the concentration of adsorbates is small. This would produce a modulated
electron concentration profile that would generate a modulated conductivity.

For illustrative purposes alone, we model the conductivity
of graphene by
\begin{equation}
 \sigma(x)=\sigma_g s(x)=\sigma_g[1-\kappa\cos(2\pi x/D)]\,,
\label{eq_sigma_x}
\end{equation}
where $\sigma_g$ is the conductivity of homogeneous graphene and $\kappa$
is related to either the strain field field or the molecular doping
concentration. Eq. (\ref{eq_sigma_x}) should be considered  a toy model.

The conductivity of graphene is a sum of two contributions: (i) a Drude term,
describing intra-band processes and (ii) a term describing inter-band transitions. At zero
temperature the optical conductivity has a simple analytical expression
\cite{nmrPRB06,falkovsky,rmp,rmpPeres,StauberGeim}.
The inter-band contribution has the form $\sigma_I=\sigma_I'+i\sigma_I''$,
with
\begin{equation}
 \sigma_I'=\sigma_0\left(
1 + \frac{1}{\pi}\arctan\frac{\hbar\omega-2\epsilon_F}{\hbar\gamma}
-
\frac{1}{\pi}\arctan\frac{\hbar\omega+2\epsilon_F}{\hbar\gamma}
\right)\,,
\end{equation}
and 
\begin{equation}
 \sigma_I''=-\sigma_0
\frac{1}{2\pi}
\ln\frac{(2\epsilon_F+\hbar\omega)^2+\hbar^2\gamma^2}
{(2\epsilon_F-\hbar\omega)^2+\hbar^2\gamma^2}\,,
\end{equation}
where $\sigma_0=\pi e^2/(2h)$.
The  Drude conductivity term is
\begin{equation}
\sigma_D =\sigma_0\frac{4\epsilon_F}{\pi}\frac{1}{\hbar\gamma-i\hbar\omega}\,,
\label{eq_sigma_xx_semiclass}
\end{equation}
where $\gamma$ is the relaxation rate and $\epsilon_F>0$ is the (local)
Fermi level position with respect to the Dirac point.
The total conductivity is
\begin{equation}
 \sigma_g=\sigma'+i\sigma''=\sigma_I'+i\sigma_I''+\sigma_D\,.
\end{equation}
\begin{figure}[!h]
 \begin{center}
 \includegraphics*[width=10cm]{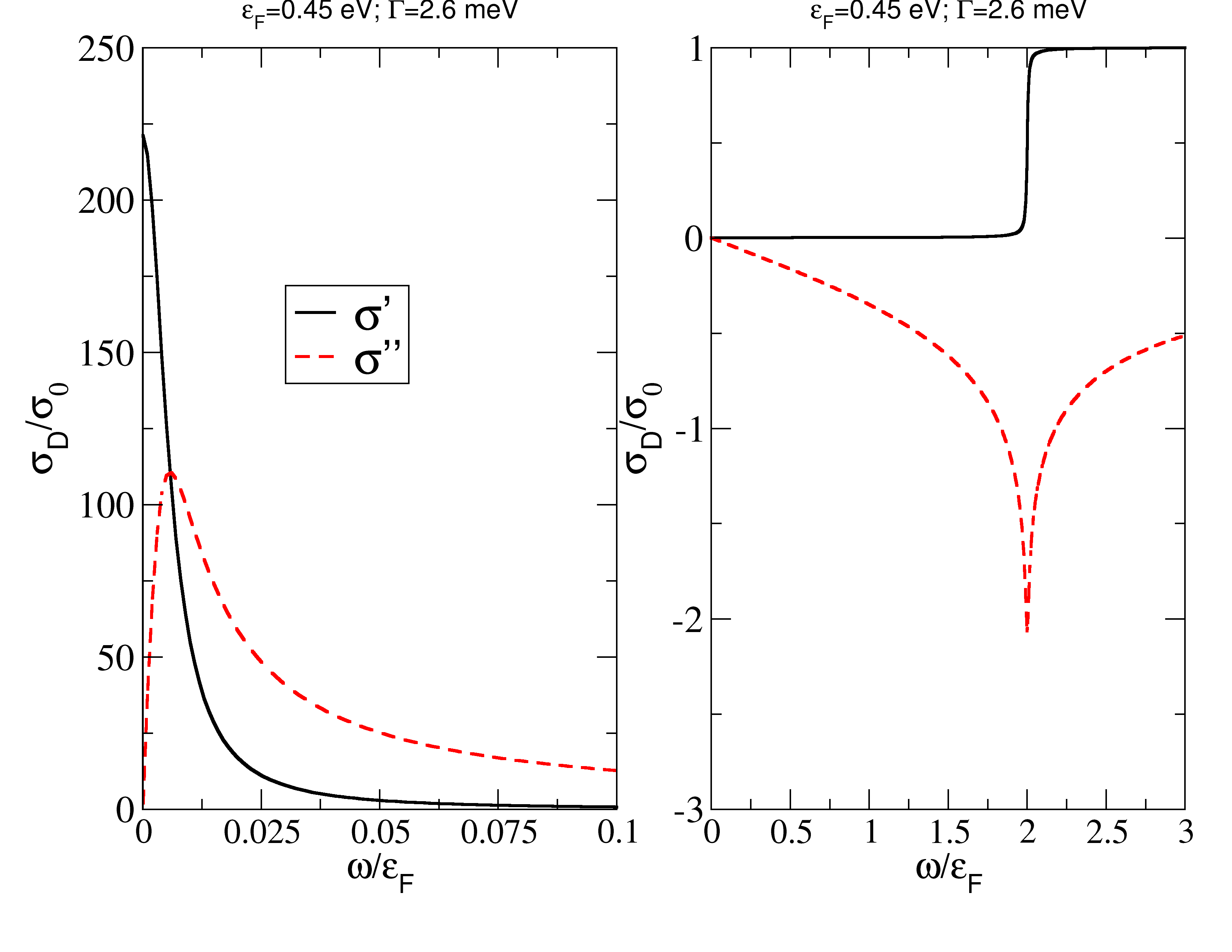}
 \end{center}
\caption{Optical conductivity of uniform
of graphene:  Drude (left) and inter-band (right) contributions.
We assume $\epsilon_F=0.45$ eV and $\Gamma=2.6 $ meV. The solid (dashed) line
stands for the real (imaginary) part of the conductivity.}
\label{fig_conductivity}
\end{figure}
In Fig. \ref{fig_conductivity} we plot the two contributions, Drude and inter-band, separately
for a given value of $\epsilon_F$ and $\Gamma=\hbar\gamma$. For heavily doped graphene
and for photon energies $\hbar\omega/\epsilon_F\lesssim0.5$ the optical response is dominated
by the Drude term. Therefore, in what follows we assume
\begin{equation}
 \sigma_g\approx\sigma_D\,,
\end{equation}
since we are  interested in the regime of frequencies $\hbar\omega/\epsilon_F\ll0.5$.
For the frequency range of interest in this work (the THz spectral range) the above 
approximation gives accurate results. For the simulations given ahead we assume graphene
sandwiched between two dielectrics of relative permittivities $\epsilon_1=3$ and 
$\epsilon_2=4$, and an average  Fermi level $\epsilon_F=$0.45 meV, as in the experiments
of Ref. \cite{LongJuPlasmonics}. In the same experiments the array of micro-ribbons has
a period of $8$ $\mu$m. In this work we choose $D=10$ $\mu$m.

\section{SPP dispersion}
\label{sec_spp_disp}
Before solving the problem of ER scattering by a modulated conductivity, 
we overview some key concepts  on surface
 plasmons-polaritons (SPPs) dispersion relation. 
SPPs are hybridized
modes of electromagnetic radiation and free electrons of a conductor. They
 propagate along the interface of a dielectric and a conductor,  and
decay exponentially away from the interface. In the case of graphene,  SPPs
propagate along the graphene sheet. The SPPs have  $p-$polarization 
(TM-wave) and their  spectrum  is given by \cite{Falko,ZieglerPRL,Jablan}:
\begin{equation}
 1 + \frac{\kappa_1\epsilon_2}{\epsilon_1\kappa_2}
+i\sigma_g\frac{\kappa_1}{\omega\epsilon_1}=0\,,
\label{eq_Wk1k2}
\end{equation}
and 
\begin{equation}
\kappa_m^2=q^2-\omega^2/v_m^2\,, 
\end{equation}
where $m=1,2$ labels the media 1 (above) and 2 (below) the graphene sheet,
$v_m=\sqrt{1/(\mu_m\epsilon_m)}$ is the velocity of light in medium $m$, $q$
is the propagation wave number along the interface, and $\mu_m$ is the
permeability of medium $m$ (we shall assume that $\mu_m=\mu_0$, the value in the vacuum).
When the two media are equal,  we obtain a simpler relation for the spectrum
of the SPPs
\begin{equation}
 1+i\frac{\sigma_g}{2\omega\epsilon}\sqrt{q^2-\omega^2\epsilon_1\mu_1}=0\,.
\label{eq_Wk1}
\end{equation}
The wave number $\kappa_m$ gives the degree of localization of the SPP. 
This is defined as the ratio (assuming graphene in vacuum) \cite{Koppens2011}
\begin{equation}
 \frac{\omega}{c\kappa_1}\approx2\alpha_f\frac{\epsilon_F}{\hbar\omega}
\end{equation}
where $\alpha_f$ is the fine structure constant in free space and
 we have used Drude's formula for the conductivity, 
and took the limit $\Gamma\ll\hbar\omega$. If we plug in typical
numbers (see section ahead) we obtain
\begin{equation}
\frac{\omega}{c\kappa_1}  \approx \frac{2}{137}\frac{0.45 \mbox{ eV}}{15 \mbox{ meV}}=0.4\,,
\end{equation}
meaning a confinement size smaller than the wavelength
of light in free space. Depending on the value of the 
Fermi energy and on the frequency of interest the degree of localization
can be much smaller than 1.


In general, Eqs. (\ref{eq_Wk1k2}) and (\ref{eq_Wk1}) cannot be solved analytically.
However, approximating $\sigma_g$  by its imaginary part (dispersive conductor
approximation)
\begin{equation}
 \sigma_g\approx i\sigma_0\frac{4\epsilon_F}{\pi\hbar\omega}\,,
\label{eq_S_diel}
\end{equation}
and taking (for simplicity) $\epsilon_1=\epsilon_2=\epsilon_0$, the vacuum value
for the permittivity,
we obtain
\begin{equation}
 q^2=\frac{\omega^2}{c^2}\left[
1+\left(
\frac{\hbar\omega}{2\alpha_f\epsilon_F}
\right)^2
\right]\,,
\end{equation}
or
\begin{equation}
\omega^2=2\left(
\frac{\alpha_f \epsilon_F}{\hbar}\right)^2\left[
\sqrt{1+\left(\frac{\hbar c q}{\alpha_f\epsilon_F}\right)^2}-1
\right]\,.
\end{equation}
In the limit $2\hbar c q/(\alpha_f\epsilon_F)\gg1$ (non-retarded approximation)
we obtain for the dispersion relation of the SPP the simple formula
\begin{equation}
 \hbar\omega = \sqrt{2\alpha_f\epsilon_F\hbar c q}\,.
\end{equation}
In the limit $q\rightarrow0$ we find that the 
SPP dispersion relation tends to $\omega\rightarrow qc$. 
For the case where $\epsilon_1\ne\epsilon_2$, the non-retarded
approximation yields

\begin{equation}
 \hbar\omega=\sqrt{2\bar\alpha_f \epsilon_F \hbar cq}\,,
\label{eq_SPPW}
\end{equation}
where 
\begin{equation}
\bar\alpha_f = \frac{e^2}{4\pi\bar\epsilon\hbar c}\,, 
\end{equation}
with $\bar\epsilon=(\epsilon_1+\epsilon_2)/2$. We note that the behaviour of $\omega(q)$
as $\sim\sqrt{q}$ cannot be trusted at small $q$, because for those values of 
$q$ the non-retarded
approximation is violated. In this case a full numerical solution of Eq. (\ref{eq_Wk1k2})
is necessary.
In what concerns the problem of ER scattering discussed ahead we will always be in the
regime where the non-retarded
approximation holds.

\begin{figure}[!ht]
 \begin{center}
 \includegraphics*[width=9cm]{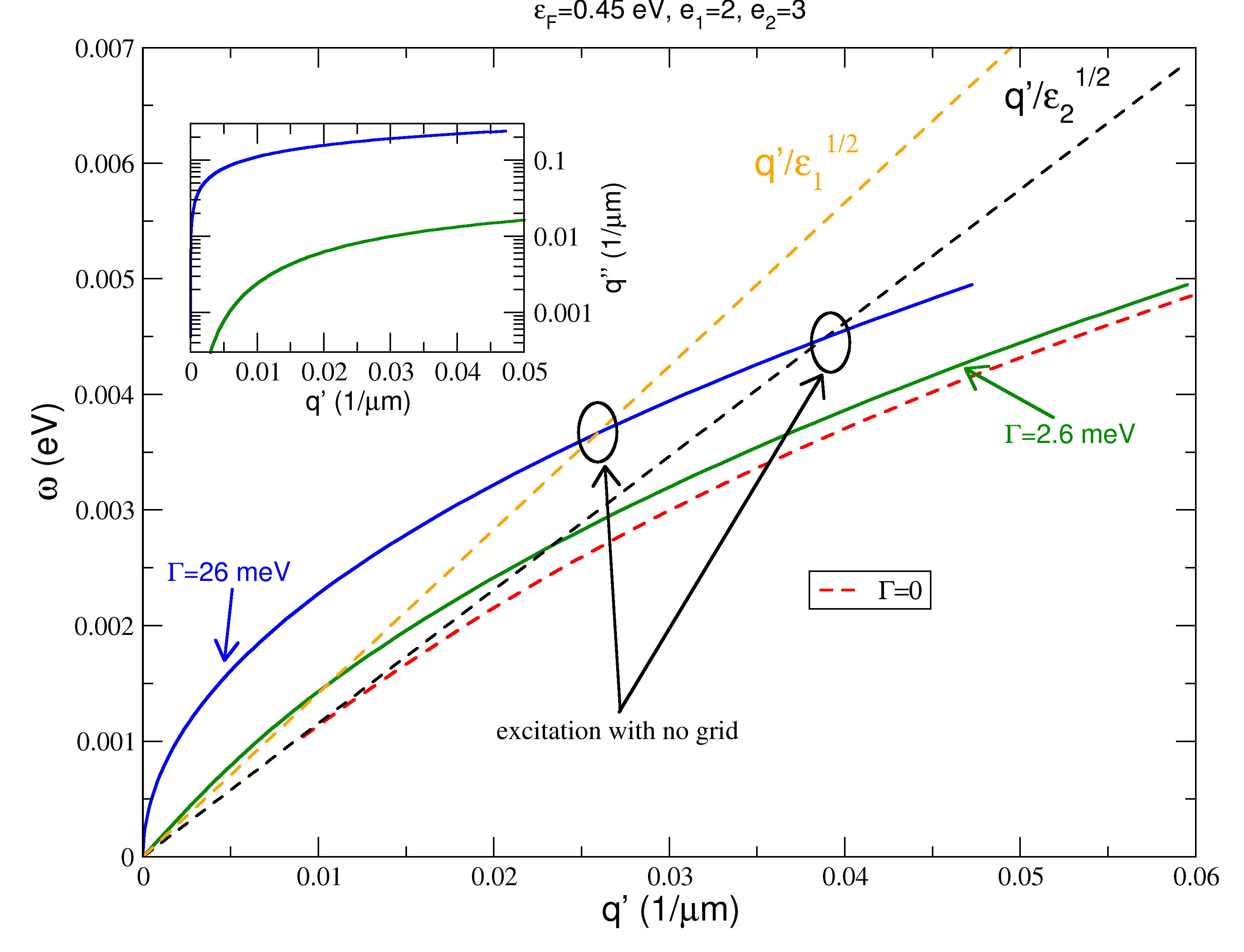}
 \end{center}
\caption{Plasmons-polaritons dispersion curves considering the effect of the 
broadening. The full conductivity $\sigma_D$ is included. The inset shows the dependence of 
the imaginary part of $q$, $q''$, as function of the real part of $q$, $q'$.}
\label{fig_broadening}
\end{figure}

\begin{figure}[!ht]
 \begin{center}
 \includegraphics*[width=9cm]{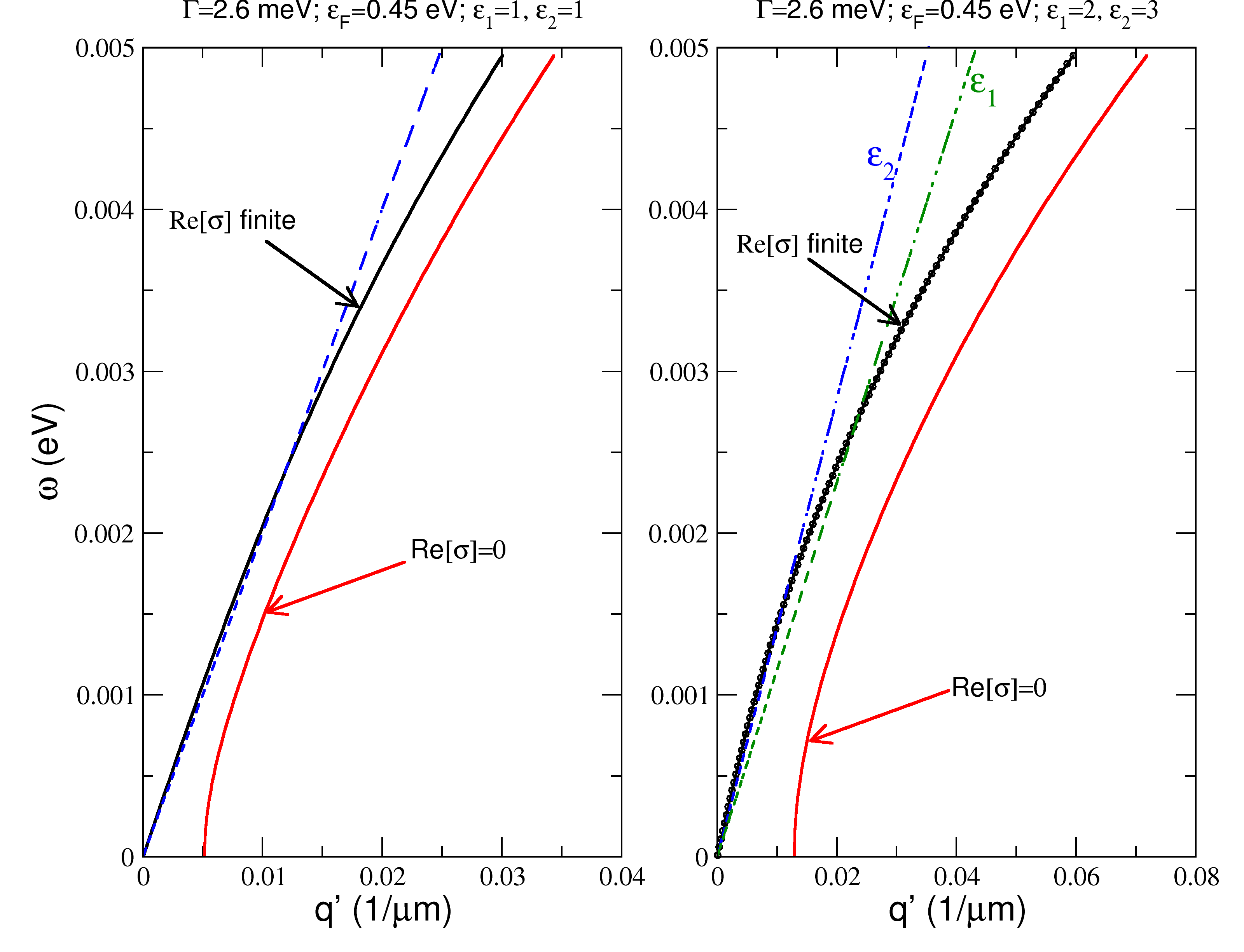}
 \end{center}
\caption{Plasmons-polaritons dispersion curves considering that 
the real part of the 
conductivity  either finite or zero, and
taking the cases of equal and different dielectrics sandwiching graphene.
The straight lines marked $\epsilon_1$ and $\epsilon_2$ are the
light dispersion $\hbar\omega=\hbar c q/\sqrt{\epsilon_1}$ and 
$\hbar\omega=\hbar c q/\sqrt{\epsilon_2}$ in media 1 and 2, respectively.}
\label{fig_real}
\end{figure}

In Figs. \ref{fig_broadening} and \ref{fig_real} we present the 
low momentum behaviour of the SPPs spectrum 
comparing the numerical solution of Eq. (\ref{eq_Wk1k2})
for different levels of approximation to the optical conductivity and considering
the role of the damping $\Gamma=\hbar\gamma$.
In Fig. \ref{fig_broadening} the effect of the damping is studied, keeping
the full conductivity $\sigma_g=\sigma_D$. In this case the momentum $q$ becomes
a complex number, $q=q'+iq''$, where $q''$ describes the decay of the SPPs as it 
propagates in space along the graphene sheet. From this figure we see that the effect of the 
increase of the damping is two-fold: it shifts the SPPs dispersion relation
toward higher energies and enhances the value of $q''$ (as expected).
If one wants to have long propagation lengths for the SPPs 
$\Gamma$ must be as small as possible. In the same figure we also 
represent the light-lines $\hbar c q/\sqrt{\epsilon_1}$ and $\hbar c q/\sqrt{\epsilon_2}$
which are the dispersion relations of a photon propagating in media 1 and 2, respectively.
At grazing incidence it is possible to excite SPPs in graphene 
with a momentum $q'$ corresponding to the point where the light line
intercepts the dispersion curve of the SPPs. This occurs, unfortunately, at
very low energies, $\nu=\omega/(2\pi)<1$ THz, which corresponds to $4.2$ meV.

In Fig. \ref{fig_real} we present the effect of neglecting the real part of the 
optical conductivity. The central feature is the vanishing of the dispersion curve for 
a finite value of $q'$. This is a spurious result: when we use the full Drude conductivity
the dispersion vanishes at zero $q'$. Changing the dielectric constants of the 
surrounding media changes the value of $q'$ for the same frequency, 
as can be seen comparing the left
and right panels of Fig.  \ref{fig_real}.

\begin{figure}[!ht]
 \begin{center}
 \includegraphics*[width=9cm]{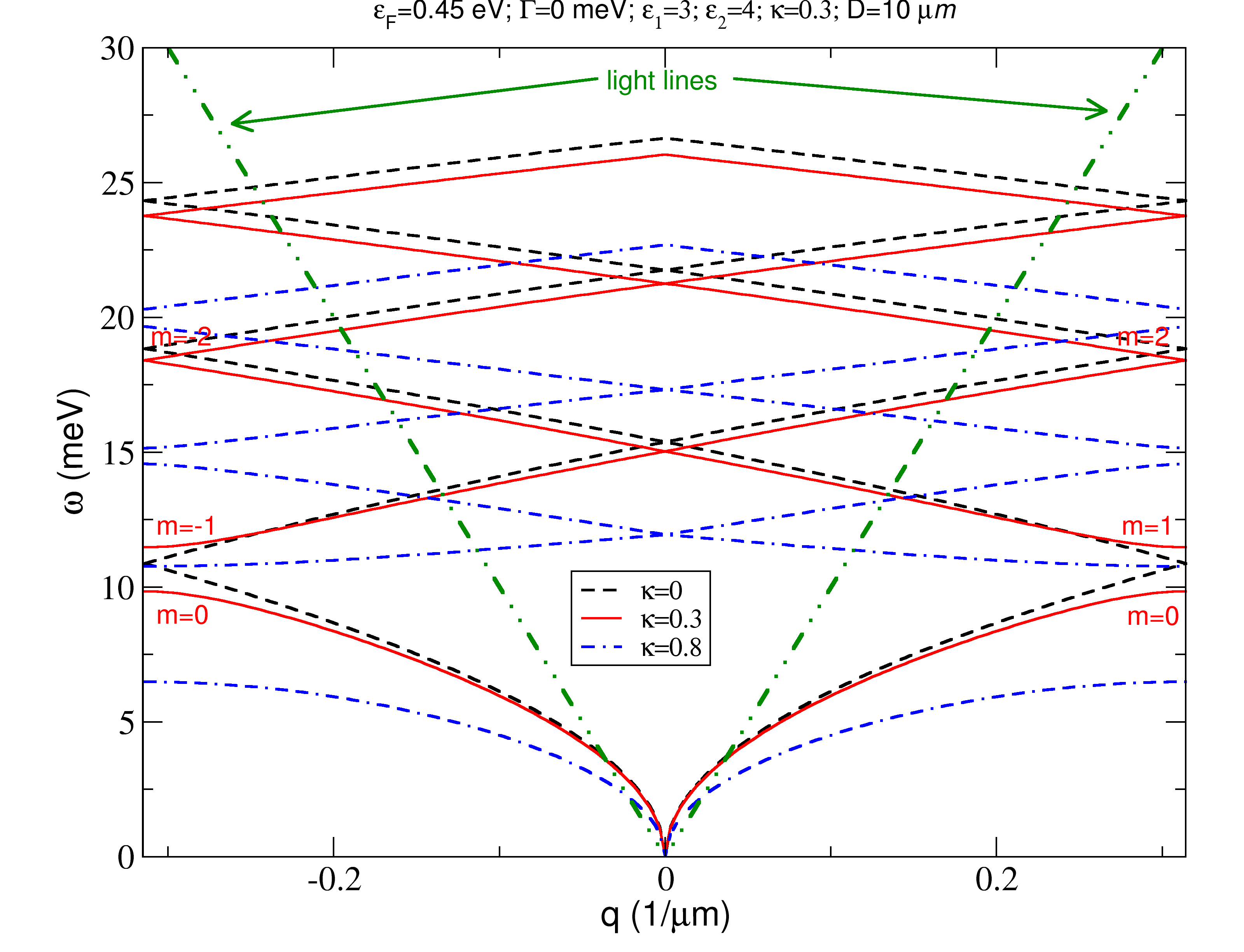}
 \end{center}
\caption{Band structure of the SPPs dispersion curves for the cosine
modulation of the conductivity given by Eq. (\ref{eq_sigma_x}).
Note that only the first six bands are shown. 
The light-cone, Eq. (\ref{eq_light_line}), at grazing incidence
($\theta=\pi/2$) is
represented by the dashed-dotted straight lines.}
\label{fig_BS_cosine_profile}
\end{figure}

When a periodic modulation of the conductivity exists, the SPPs 
develop a band structure depicted in Fig. \ref{fig_BS_cosine_profile}.
The black dashed lines represent the folding of the dispersion curve (\ref{eq_SPPW})
into the first Brillouin zone, for vanishing modulation of the
conductivity ($\kappa=0$). For finite modulation of $\sigma(x)$ 
($\kappa\ne0$) gaps develop at the edges of the Brillouin zone.
The band structure represented in Fig. \ref{fig_BS_cosine_profile} was computed
in the non-retarded approximation, considering only the imaginary part
of the conductivity (meaning that $q=q'$ and $q''=0$), and taking $\Gamma=0$.
As such, the behaviour of the first band close to $q=0$ is not accurate. In this
approximation, there are no gaps at the center of the zone. 
(The approximations used allows to transform a non-linear eigen-problem into a linear one.)
The region 
of the band structure
between the 
two straight lines (see Fig. \ref{fig_BS_cosine_profile}) can be 
directly excited by light impinging on graphene
without the aid of the ATR technique. If ER impinges at an angle
$\theta$, the momentum along the graphene sheet is $k_1\sin\theta=\omega/v_1\sin\theta$
and then a SPP with momentum $q=\omega/v_1\sin\theta$ can be created. This corresponds
to a frequency of 
\begin{equation}
 \omega=\frac{qc}{\sqrt{\epsilon_1}\sin\theta}\,.
\label{eq_light_line}
\end{equation}
When the straight line (\ref{eq_light_line}) intercepts one of the upper bands in the 
Brillouin zone a SPP can be excited. If the modulation in the conductivity is weak 
(as, for example, is the case $\kappa=0.3$; see Fig. \ref{fig_BS_cosine_profile}) 
the energy of the excited SPP is roughly
given by  
\begin{equation}
 \omega_{\mbox{\tiny SSP}}\simeq\sqrt{2\bar\alpha_f\epsilon_F \hbar c
\vert k_1\sin\theta-mG}\vert\,,
\end{equation}
where $G=2\pi/D$ and $m$ an integer.

To conclude this section, we highlight the result  we use later
in the interpretation of ER scattering: for large $q'$ the dispersion relation
is approximately given by Eq. (\ref{eq_SPPW}) and at high energies SPPs
cannot be excited by shining light on homogeneous graphene, even at grazing
incidence. When 
the optical conductivity
is modulated periodically
 the system develops a band structure for the SPPs dispersion
relation
and their direct excitation  becomes possible.

\section{Electromagnetic radiation scattering: formalism}
%
%
As discussed in Sec. \ref{sec_mod_conductivity} a conductive surface bearing a modulated 
conductivity supports SPPs which can be directly excited by light impinging
on the conductor's surface, that is, without the need of a grating.
We develop now the formalism needed to study, in a quantitative way, ER scattering by a 
modulated conductivity.

Since the conductivity is a periodic function
we can write it as a Fourier
series ($G=2\pi/D$):
\begin{equation}
 \sigma(x)=\sum_{m=-\infty}^{\infty}e^{iGmx}\sigma_{mG}\,,
\end{equation}
where 
\begin{equation}
 \sigma_{mG}\equiv\sigma(mG)=\frac{1}{D}\int_0^D\sigma(x)e^{-imGx}dx\,.
\end{equation} 
Considering  $p-$polarization,
we write the electromagnetic fields as series of Bloch waves

\begin{eqnarray}
 B_{y}^\lambda(x,z)&=&\delta_{1,\lambda}b_ie^{izk_0\sin\theta }e^{ixk_0\cos\theta }+
\sum_{n=-\infty}^\infty b^{(\lambda)}_{k-nG}e^{i(k_0\sin\theta-nG)z}e^{-q^{(\lambda)}_n\vert x\vert}\,,\nonumber\\
 E_{z}^\lambda(x,z)&=&\delta_{1,\lambda}e_{z;i}e^{izk_0\sin\theta }e^{ixk_0\cos\theta }+
\sum_{n=-\infty}^\infty e^{(\lambda)}_{z;k-nG}e^{i(k_0\sin\theta-nG)z}e^{-q^{(\lambda)}_n\vert x\vert}\,,\nonumber\\
 E_{x}^\lambda(x,z)&=&\delta_{1,\lambda}e_{x;i}e^{izk_0\sin\theta }e^{ixk_0\cos\theta }+
\sum_{n=-\infty}^\infty e^{(\lambda)}_{x;k-nG}e^{i(k_0\sin\theta-nG)z}e^{-q^{(\lambda)}_n\vert x\vert}\,,
\end{eqnarray}
where $\lambda=1,2$ labels the two cladding media.
The relations between the amplitudes of the fields follow from Maxwell's
equations:
\begin{enumerate}
 \item incoming field ($x<0$):
\begin{eqnarray}
e_{x;i} &=& -\frac{\sin\theta}{\cos\theta}e_{z;i}\,,\\
b_i &=& \frac{\mu_0\epsilon_1\omega}{k_0\cos\theta}e_{z;i}\,. 
\label{eq_rel_1}
\end{eqnarray}
\item reflected field ($x<0$):
\begin{eqnarray}
e^{(1)}_{x;k_n} &=& -i\frac{k_0\sin\theta}{q_n^{(1)}}e^{(1)}_{z;k_n}\,,\\
b^{(1)}_{k_n} &=& -i\frac{\mu_0\epsilon_1\omega}{q_n^{(1)}}e^{(1)}_{z;k_n}\,. 
\label{eq_rel_2}
\end{eqnarray}
\item transmitted field ($x>0$):
\begin{eqnarray}
e^{(2)}_{x;k_n} &=&  i\frac{k_0\sin\theta}{q_n^{(2)}}e^{(2)}_{z;k_n}\,,\\
b^{(2)}_{k_n} &=&      i\frac{\mu_0\epsilon_2\omega}{q_n^{(2)}}e^{(2)}_{z;k_n}\,. 
\label{eq_rel_3}
\end{eqnarray}
\end{enumerate}

The boundary condition $E_z^1=E_z^2$ implies:
\begin{eqnarray}
 e_{z;i} + e^{(1)}_{z;k_0} &=& e^{(2)}_{z;k_0}\,,\\
   e^{(1)}_{z;k_n} &=& e^{(2)}_{z;k_n}\hspace{0.3cm} \wedge \hspace{0.3cm} n\ne0 \,.
\end{eqnarray}
The second boundary condition, $B_y^1-B_y^2=-\mu_0\sigma E_z^1$, implies 
\begin{eqnarray}
 b_{i} + b^{(1)}_{k_0} &=& b^{(2)}_{k_0} -\mu_0\sum_m\sigma(mG)e^{(2)}_{z;k_m}\,,\\
   b^{(1)}_{k_n} &=& b^{(2)}_{k_n} -\mu_0\sum_m\sigma(mG)e^{(2)}_{z;k_{m+n}}
\hspace{0.3cm} \wedge \hspace{0.3cm} n\ne0 \,.
\end{eqnarray}
Using the relations between fields, Eqs. (\ref{eq_rel_1})-(\ref{eq_rel_3}),
the set of boundary conditions reduces to 
\begin{eqnarray}
\label{eq_linear_a}
 \frac{\epsilon_1\omega}{q^{(1)}_0}e^{(2)}_{z;k_0}+ 
\frac{\epsilon_2\omega}{q^{(2)}_0}e^{(2)}_{z;k_0}
+i\sum_m\sigma(mG)e^{(2)}_{z;k_m}&=&\frac{2i\epsilon_1\omega}{k_0\cos\theta}e_{i;z}\,,\\
\label{eq_linear_b}
\frac{\epsilon_1\omega}{q^{(1)}_n}e^{(2)}_{z;k_n}+ 
\frac{\epsilon_2\omega}{q^{(2)}_n}e^{(2)}_{z;k_n}
+i\sum_m\sigma(mG)e^{(2)}_{z;k_{m+n}}&=&0 \hspace{0.3cm} \wedge \hspace{0.3cm} n\ne0\,.
\end{eqnarray}
We note that $q_0^{(1)}=-ik_0\cos\theta$, with $k_0=\omega/v_1$. We further define
\begin{equation}
 q^{(\lambda)}_n=\sqrt{(k_0\sin\theta-nG)^2-\omega^2/v_\lambda^2}\,,
\label{eq_q}
\end{equation}
for a positive argument of the square root (evanescent waves);
if the argument of the square  root in Eq. (\ref{eq_q}) is negative 
(propagating waves)
$q^{(\lambda)}_n$
is written as 
\begin{equation}
 q^{(\lambda)}_n = -i\sqrt{\omega^2/v_\lambda^2-(k_0\sin\theta-nG)^2}\,.
\end{equation}
The choice of sign for the square root is dictated by physical reasons:
 reflected waves for $x<0$ and transmitted waves for $x>0$.
The ``pseudo-reflectance amplitude'' of the mode $n=0$ is defined as
\begin{equation}
 r_{z;0}=\frac{e^{(1)}_{z;k_0}}{e_{z;i}}=\frac{e^{(2)}_{z;k_0}}{e_{z;i}}-1\,,
\end{equation}
and that of a mode $n\neq0$
is given by
\begin{equation}
 r_{z;n}=\frac{e^{(1)}_{z;k_n}}{e_{z;i}}=\frac{e^{(2)}_{z;k_n}}{e_{z;i}}\,.
\end{equation}
The reflectance of the order $n\ne0$ (for propagating modes) reads
\begin{eqnarray}
 R_n
=i\frac{k_0\cos\theta}{q^{(1)}_n}\left\vert\frac{e^{(1)}_{z,k_n}}{e_{z,i}}\right\vert^2\,.
\end{eqnarray}
The reflectance and the transmittance 
of the specular ($n=0$) mode are given by
\begin{equation}
 {\cal R}=\vert r_0\vert^2\,,
\end{equation}
and
\begin{equation}
 {\cal T} = 
\left\vert \frac{e^{(2)}_{z,k_0}}{e_{z,i}}\right\vert^2\frac{\epsilon_2}{\epsilon_1}
\frac{\cos\theta}{\sqrt{\epsilon_2/\epsilon_1-\sin^2\theta}}\,,
\end{equation}
respectively.
The last expression is valid for $\epsilon_2>\epsilon_1$
or for $\theta<\theta_c=\arcsin(\epsilon_2/\epsilon_1)$, 
the critical angle for total reflection.
In what follows only the the specular mode is propagating 
(all the remaining modes are evanescent). Then, the absorbance is simply
given by ${\cal A}=1-{\cal R}-{\cal T}$.

\section{Special limits}

Here we show that the linear system of Eqs. (\ref{eq_linear_a}) and (\ref{eq_linear_b})
reduces to well known formulae when the conductivity of graphene is either zero
or finite and homogeneous. If the conductivity of graphene vanishes,
 we simply have 
an interface between two different dielectrics. In this case Eqs. (\ref{eq_linear_a})
and (\ref{eq_linear_b}) give $e^{(2)}_{z;k_n}=0$ ($n\ne0$)and
\begin{equation}
 \frac{e^{(2)}_{z;k_0}}{e_{z;i}}=2
\left(
1+\sqrt{\frac{\epsilon_2}{\epsilon_1}}\frac{\cos\theta}
{\sqrt{1-\epsilon_1\sin^2\theta/\epsilon_2}}
\right)^{-1}\,.
\end{equation}
The reflectance amplitude follows from
\begin{equation}
 r_0=1-\frac{e^{(2)}_{z;k_0}}{e_{z;i}}=
\frac{\sqrt{\epsilon_2/\epsilon_1}\cos\theta - \sqrt{1-\epsilon_1\sin^2\theta/\epsilon_2} }
{\sqrt{\epsilon_2/\epsilon_1}\cos\theta + \sqrt{1-\epsilon_1\sin^2\theta/\epsilon_2}}\,,
\end{equation}
which reproduces the well known result from elementary optics.

Another particular limit is obtained when $\sigma(x)$ is 
finite and homogeneous. In this case, only the 
Fourier component $m=0$ of $\sigma(x)$ exists. Thus, we obtain
\begin{equation}
 \frac{e^{(2)}_{z;k_0}}{e_{z;i}} 
\left(
\frac{1}{\cos\theta}+\frac{\epsilon_2/\epsilon_1}{\sqrt{\epsilon_2/\epsilon_1-\sin^2\theta}}
+\frac{\sigma_D}{\epsilon_1v_1}\right)=\frac{2}{\cos\theta}\,.
\end{equation}
For $\epsilon_1=\epsilon_2$ we obtain the well known result for the 
transmittance of graphene (for a TM wave) \cite{StauberGeim}
\begin{equation}
  {\cal T}=\left\vert\frac{2}{2+\sigma_D\cos\theta/\epsilon_1v_1}\right\vert^2\,.
\end{equation}
\section{Electromagnetic radiation scattering: results}
%
%
When the optical conductivity of graphene has the form (\ref{eq_sigma_x}), the 
Fourier transform of $s(x)$ reads:
\begin{eqnarray}
 s(0)&=&1\,,\\
 s(mG)&=&-\frac{\kappa}{2}(\delta_{m,1}+\delta_{m,-1})\,.
\end{eqnarray}
The linear system defined by Eqs. (\ref{eq_linear_a}) and (\ref{eq_linear_b})
is solved numerically and the sums over $m$ 
are cutoff at $m=-N,\ldots,0,\ldots,N$; the numerical solution rapidly converges
with $N$. 

\begin{figure}[!htb]
 \begin{center}
 \includegraphics*[width=9cm]{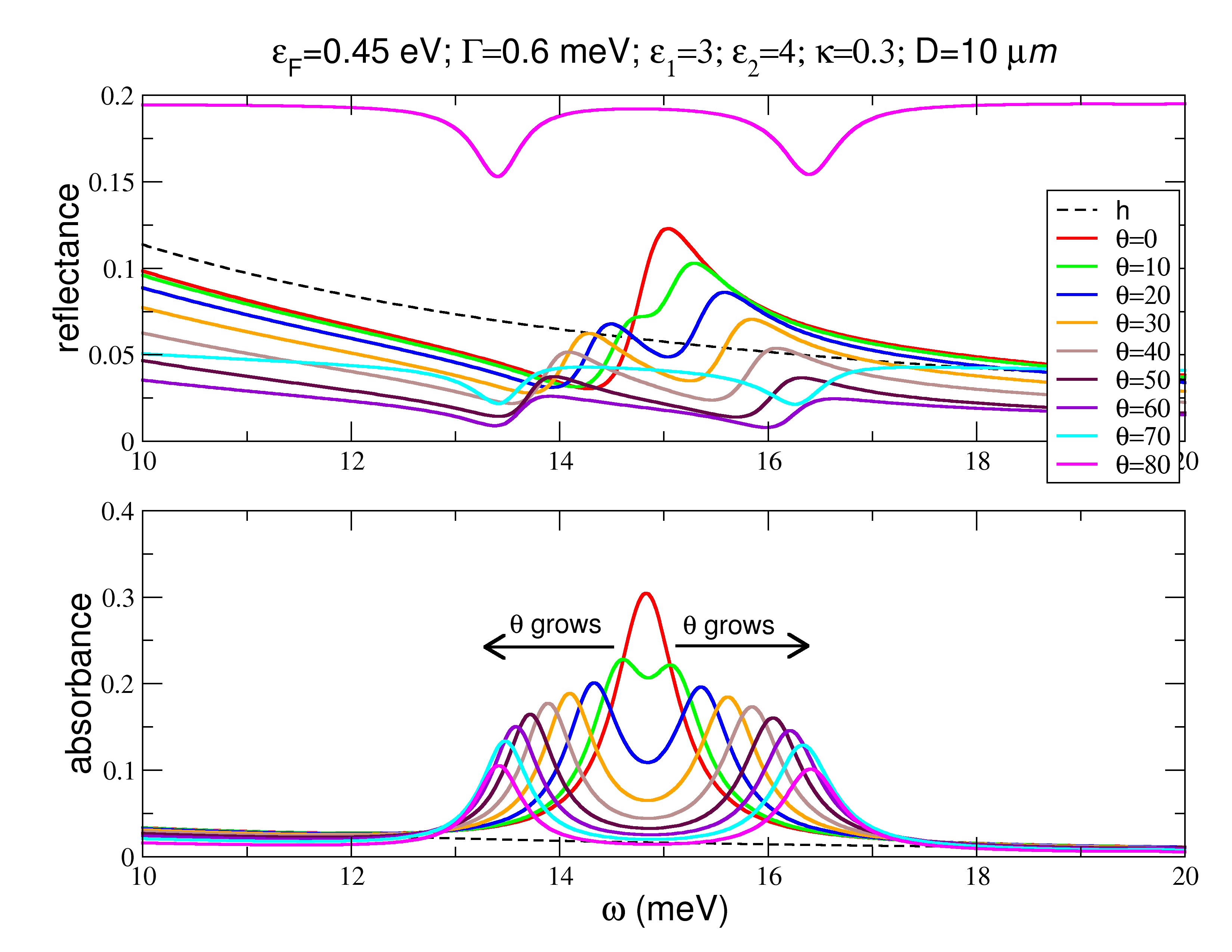} 
 \end{center}
\caption{Absorbance of a graphene sheet with a cosine-modulated conductivity.
Reflectance (top) and absorbance (bottom) as function of the energy for different
incoming angles.
The parameters are $\epsilon_F=0.452$ eV, $D=10$ $\mu$m, 
$\epsilon_1=3$, $\epsilon_2=4$,  $\hbar\Gamma=0.6$ meV, and $\kappa=0.3$.
The Brewster angle for the two dielectrics is 49.1$^o$.}
\label{fig_modulated_sigma}
\end{figure}

Results for the specular reflectance, ${\cal R}$, and for the 
absorbance, ${\cal A}$, are given in Fig. \ref{fig_modulated_sigma}. 
As discussed in Sec. \ref{sec_spp_disp}, it is not possible to excite
SPPs in a homogeneous system because of the mismatch of the momentum of
the incoming ER and that of SPP for the same frequency $\omega$.
For a modulated conductivity the momentum of the SPPs is conserved up to
a reciprocal lattice vector $mG$, with $m=\pm1,\pm2,\ldots$, that is
\begin{equation}
 q_{\mbox{\tiny SPP}}=\vert k_1\sin\theta - mG\vert\,.
\label{eq_q_SPP}
\end{equation}
In this case, even for normal incidence, it is possible
to excite  SPPs. We stress that in the present case it is the periodicity
of the conductivity and not an external grating leading to the 
validity of Eq. (\ref{eq_q_SPP}). The excitation of the SPPs at normal
incidence is clearly seen in   Fig. \ref{fig_modulated_sigma}.
In this figure both the reflectance and the absorbance are shown and 
a clear signature of SPPs excitation is present in both plots.
The dashed black curve represents the behaviour of the system for a 
homogeneous conductivity and impinging ER at normal incidence;
clearly 
the curve is featureless. For the inhomogeneous case  a large
enhancement of the absorbance is seen around the energy given by Eq. 
(\ref{eq_SPPW}) with $q=2\pi/D$. The position of the peak does not
coincide exactly with the number given by Eq. (\ref{eq_SPPW}) because
this equation is not sensitive to  details of the band structure. 
From Fig. \ref{fig_BS_cosine_profile}
we can see that the bands for $m=\pm1$ at the zone center have an energy
of about 15 meV, coinciding with the energy for which the reflectance curve
has a maximum for $\theta=0$. 
As the  angle of the incident beam approaches the  
Brewster angle 
of the two media the reflectivity decreases
substantially. The Brewster angle of the system is not given exactly
by the well known formula $\Theta_B=\arctan\sqrt{\epsilon_2/\epsilon_1}$
due to the presence of graphene.
For incidence angles above the Brewster angle the
reflectance develops two dips and can be larger than it would be for $\theta=0$
(see curve for $\theta=80^o$).

When the incoming beam deviates from 
normal incidence (i. e., $\theta\ne 0$), there is a peak splitting 
both in the reflectance and in the absorbance curves. We would like to understand 
in qualitative terms the
behaviour of the peak slitting as function of the incoming angle $\theta$.
Using Eq. (\ref{eq_q_SPP}) in Eq. (\ref{eq_SPPW}) yields
 \begin{equation}
 \hbar\omega=\sqrt{2\bar\alpha_f \epsilon_F \hbar c\vert k_1\sin\theta - mG \vert}\,,
\label{eq_WAbsspp}
\end{equation}
with $k_1=\omega/v_1$.
Clearly, when $m<0$ and $\theta$ increases the frequency shifts toward
higher energies. On the other hand, when  $m>0$ the energy 
decreases as $\theta$ increases. This behavior can be understood from the 
analysis of Fig. \ref{fig_BS_cosine_profile}. For $\theta=0$ the
light line, Eq. (\ref{eq_light_line}), is vertical and 
touches the second band at the center of the zone ($q=0$)
where the branches associated with  $m=1$ and $m=-1$ touch each other.
As $\theta$ grows, the slope of light line decreases and the branches
with both $m=1$ and $m=-1$ are no longer degenerate.

It is possible to obtain a simple
analytical expression for $\omega(\theta)$ by solving Eq. (\ref{eq_WAbsspp})
for $\hbar\omega$. The final result is 
\begin{eqnarray}
\label{eq_branch_minus_plus}
 \hbar\omega &=& \pm\frac{\sin\theta}{2a}+\frac{1}{2a}\sqrt{\sin^2\theta+4ac_1}\,,
\end{eqnarray}
for $m=-1$ and $m=1$, respectively, and the constants $a$ and $c_1$ are given by
\begin{eqnarray}
 a&=&\frac{1}{4\alpha_f\epsilon_F}\frac{\epsilon_1+\epsilon_2}{\sqrt{\epsilon_1}}\,,\\
c_m&=&m\frac{2\pi}{D}\frac{c\hbar}{\sqrt{\epsilon_1}}\,.
\end{eqnarray}
\begin{figure}[!htb]
 \begin{center}
 \includegraphics*[width=9cm]{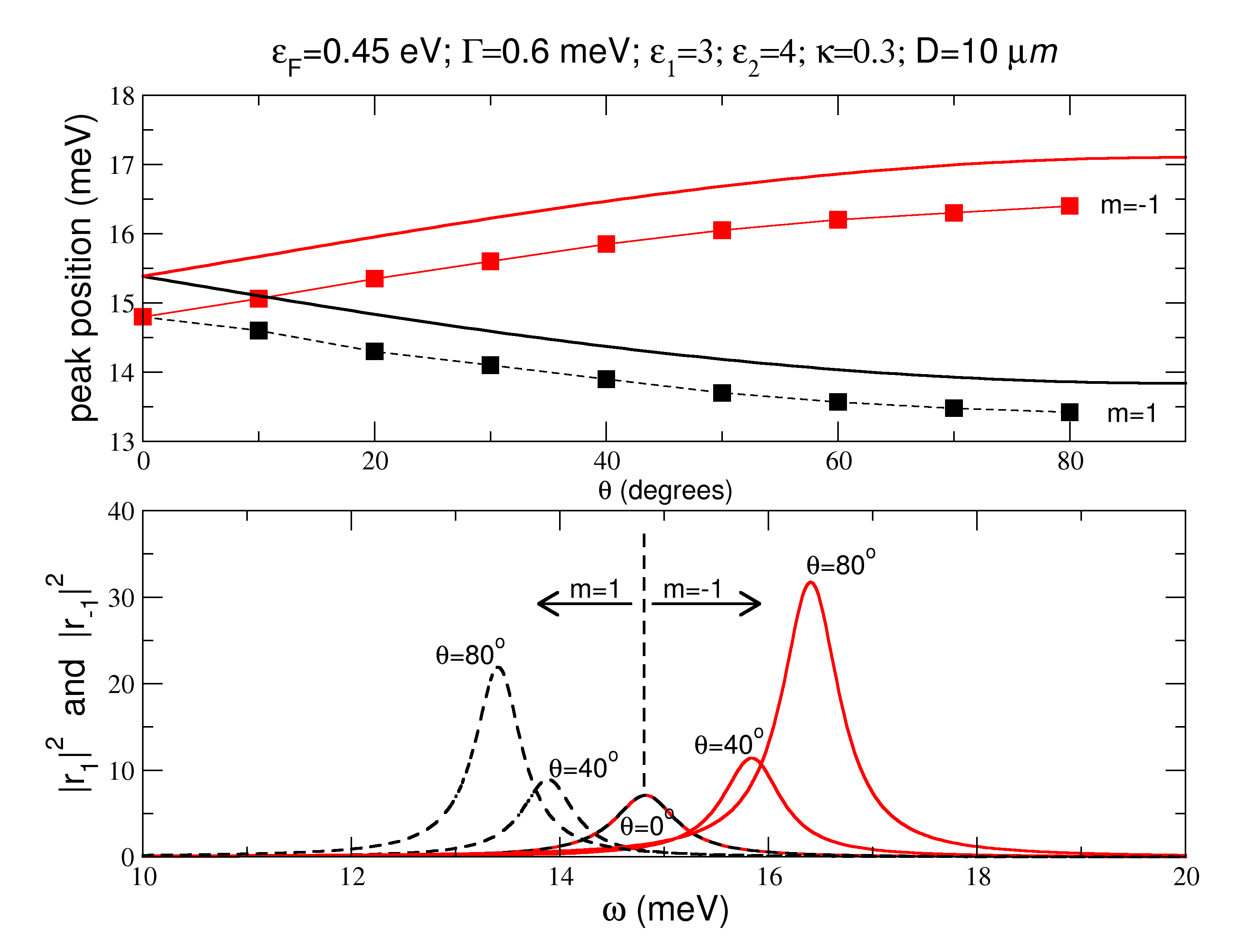} 
 \end{center}
\caption{Position of the peak of the absorbance as function of
the incoming angle.
Top: dependence of the  energy for maximum absorbance
on the angle $\theta$. The lines with squares are obtained from the bottom panel
of Fig. \ref{fig_modulated_sigma}; 
the solid lines are the two branches of Eq. 
(\ref{eq_branch_minus_plus}).
The bottom panel shows the squared absolute value of the amplitudes  of the modes associated
with the SPPs of momentum $\pm G$.
The parameters are $\epsilon_F=0.45$ eV, $D=10$ $\mu$m, 
$\epsilon_1=3$, $\epsilon_2=4$,  $\hbar\Gamma=0.6$ meV, and $\kappa=0.3$.}
\label{fig_peak_position}
\end{figure}

In Fig. \ref{fig_peak_position} we plot the two branches of Eq. 
(\ref{eq_branch_minus_plus}) and compare it with the position of the 
peaks for the absorbance obtained from Fig. \ref{fig_modulated_sigma}; the agreement
is qualitative. It cannot be quantitative because: (i) Eq. (\ref{eq_branch_minus_plus})
is derived from a kinematic argument and, therefore, misses  the dependence on
$\kappa$; (ii) we have used the non-retarded approximation. 
However, for small $\kappa$ the agreement is quite good. Indeed,
 if we had shifted the solid curves in  Fig. \ref{fig_peak_position} by a constant,
they would overlap  the points (solid squares)
obtained from Fig. \ref{fig_modulated_sigma}.
In the bottom panel of Fig. \ref{fig_peak_position} we plot
$\vert r_{\pm1}\vert^2$ for different $\theta$ as a function of the energy.
The energy of the peaks of $\vert r_{\pm1}\vert^2$ coincides with that
of peaks in the reflectance (and absorbance). This shows that the SPPs
associated with the momenta $\pm 2\pi/D$ are the ones contributing to the 
anomalies in the reflectance and absorbance spectra.

\section{Conclusions}
%
%
We have shown that a modulated conductivity in an otherwise flat graphene sheet
gives rise to a SPPs band structure with gaps opening at the edges of the 
Brillouin zone. Within the non-retarded approximation, no gaps open at the center
of the zone. This band structure was later used to understand the qualitative
behaviour of ER scattering from a flat graphene sheet with a modulated
conductivity. The modulation allows for the SPPs be efficiently excited without
using the ATR technique. The properties of the studied system resemble those of
a photonic crystal. This analogy will be explored in a future work.

If in addition to the modulated conductivity, corrugation is induced
in the graphene sheet the reflectance and absorbance spectra are predicted to have
a richer structure due to two competing mechanisms: the inhomogeneity and the 
presence of the groves.
Finally we note that the case where  graphene is cut into micro-ribbons
is qualitatively different from the one discussed here, because the graphene sheet
is, in that case, discontinuous.

\ack
NMRP and AF thank useful discussions with Antonio Castro Neto and Kian Ping Loh.
NMRP thanks  the hospitality of and funding from the National University of Singapore (NUS),
where this work was started. MIV thanks the hospitality of the {\it Graphene Centre} 
at NUS. The work was partially supported by
the Portuguese Foundation for Science and Technology.

\section*{References}

\providecommand{\newblock}{}

\end{document}